\documentclass[aps,prd,preprint,groupedaddress,nofootinbib]{revtex4}

\newif\ifpdf
\ifx\pdfoutput\undefined
	\pdffalse
\else
	\pdfoutput=1
	\pdftrue
\fi

\usepackage{cancel}
\usepackage{epsfig}
\usepackage{graphicx}

\begin{document}


\title{The Minimal Model of Radion Mediated Dark Matter}


\author{Chris Bird}
\email[]{cbird@uvic.ca}
\affiliation{Department of Physics and Astronomy, University of
Victoria, \\
 Victoria, B.C., V8P 1A1, Canada.}


\date{\today}

\begin{abstract}
Based on the results from numerous astrophysics experiments, it is currently believed that the majority of matter in the Universe is in some unknown form, known as {\it dark matter}. In the past it has been common to model dark matter as a massive particle with weak interactions - either from existing Standard Model forces or from new forces - and often as past of a more complicated theory. In this article, an alternative is presented in which dark matter is in the form of a massive particle with no explicit interactions. It is demonstrated that, in the presence of strongly warped extra dimensions the inherent gravitational interactions of such a particle are sufficient to produce the observed abundance of dark matter without violating existing experimental constraints.
\end{abstract}


\maketitle


\section{Introduction}

\par
One of the oldest and most important problems in modern cosmology is the nature of dark matter. While several astrophysics experiments have inferred its existence and even produced precise measurements of its abundance\cite{Turner:2000kq,Spergel:2003cb,Dunkley:2008ie}, at this time there are no strong indications of the precise properties of the missing matter. Although there are numerous candidates for dark matter, among the most common is the weakly interacting massive particle.

\par
At present there are many theories which claim to predict the existence of dark matter. In many of these, new physics is introduced for other purposes (such as supersymmetry or extra dimensions), while dark matter is assumed to be a side effect of the theory. However as suggested in Ref \cite{McDonald:1993ex,Burgess:2000yq,Bird:2006jd}, it is also possible that dark matter has no connection to any other theory and exists as a single new particle. For this reason, it is useful to study minimal models of dark matter in which only a minimum of new physics is introduced.

\par
In order to produce the correct dark matter abundance, the WIMPs are constrained to annihilate with a cross-section of order $<\sigma_{ann} v> \sim O(1 \; pb)$. As such the dark matter is expected to interact only through a weak interaction, either as part of the Standard Model or through a new force. In the past,  minimal models have used one or more Higgs bosons to provide the required interaction between the dark matter candidate and the Standard Model. In this article an alternative method is introduced, in which the WIMP annihilation cross-section can be made sufficiently large without requiring non-gravitational interactions.

\par
Since WIMPs cannot interact through electromagnetic or strong nuclear forces, and since interactions through weak nuclear forces are tightly constrained by experiments, it is tempting to consider WIMPs which only interact through gravity. However in general, gravity is too weak to produce the correct abundance of thermally produced dark matter. One possible exception is to produce dark matter in regions where gravity is stronger, such as in the presence of warped extra dimensions.

\section{Dark Matter \& Warped Extra Dimensions} \label{Section::EXDM}

\par
The possible existence of extra dimensions has become very popular in recent years \cite{Arkani-Hamed:1998rs,Randall:1999ee,Randall:1999vf}, with the primary motivation for such models being a resolution of the {\it hierarchy problem}. The electroweak forces have couplings of the order $O(TeV^{-1})$, while gravitational couplings are of order $M_{PL}^{-1} = \sqrt{G_N} = 0.82 \times 10^{-16} \; TeV^{-1} $. However the Standard Model cannot explain this large difference in the strengths of the forces.

\par
In these models, the Standard Model fields are trapped on a four-dimensional spacetime brane while gravity can propagate in higher dimensions as well, effectively diluting the gravitational forces. As a result, gravitational couplings can be large without violating existing experimental bounds.

\par
For the purposes of this model, all that is required of the extra dimensions is that they lead to a large gravitational coupling with a radion. As an example, in this article the Randall-Sundrum model\cite{Randall:1999ee,Randall:1999vf} will be assumed. The spacetime metric for this model is

\begin{equation}
ds_{RS}^2 = e^{- 2 k \phi |y|} \eta_{\mu \nu} dx^{\mu}dx^{\nu} - \phi(x_{\mu})^2 dy^2
\end{equation}

\noindent
where $\phi(x_{\mu})$ behaves in the same manner as a scalar field trapped on the four-dimensional brane, and is referred to as the {\it radion}.

\par
It should be noted that there are a number of possible candidates for dark matter which are naturally contained in extra dimensional models.
 For example, when the gravitational field propagates in the higher dimensions, it can only have certain energy levels or modes due to the boundary conditions on the extra dimension. Each of these modes has the same properties as a massive particle trapped on the brane, and this effective particle is referred to as a {\it Kaluza-Klein graviton} or a {\it Kaluza-Klein mode}. Another possibility is that the brane on which the SM fields are trapped can fluctuate in the higher dimensions, forming bumps in the brane. These fluctuations can also behave like particles trapped on the brane, referred to as {\it branons}. In the early Universe, the KK gravitons  and the branons can be formed both in the decay of other particles and in the annihilations of Standard Model particles. In the same manner that WIMPs freeze-out of thermal equilibrium to form a dark matter abundance, these effective particles can also freeze-out and replicate the effects of dark matter. These models have been studied extensively in Ref \cite{Servant:2002aq,Cheng:2002ej} and Ref \cite{Cembranos:2003mr}.
\par
 In this model, it is only assumed that the dark matter candidate is a new particle and not necessarily an effect of the extra dimensions. It is also assumed that this new particle accounts for the entire dark matter abundance, although it is possible that the observed abundance is a combination of WIMPs and Kaluza-Klein gravitons or branons.  

\par
In this article, a single new particle is introduced as the dark matter candidate. Since gravitons and radions naturally couple to the energy-momentum tensor, the WIMPs naturally interact with the Standard Model without requiring additional interactions. In comparison with the Minimal Model presented in Ref. \cite{Burgess:2000yq}, this model has the additional benefit of having one less parameter as the WIMP-gravity coupling is proportional to the WIMP mass instead of an arbitrary coupling constant. Although these properties are also present in models without extra dimensions, in those cases the gravitational interaction is too weak to efficiently annihilate WIMPs in the early Universe, with typical annihilation cross sections being of order $\sigma_{ann} \sim O(m_{dm}^2 M_{PL}^{-4})$. 
 Since the Planck mass is several orders of magnitude lower in the Randall-Sundrum model, the annihilation cross-section is much larger in the presence of warped extra dimensions and the WIMPs can annihilate efficiently. 

\par
For the purposes of this article, there are two minimal models. The first model is a singlet scalar WIMP, with no non-gravitational interactions, and with Lagrangian 

\begin{equation}
L = \frac{1}{2} (\partial_{\mu} S)^2 - \frac{1}{2} m_S^2 S^2
\end{equation}

\noindent
The second model is similar, except the WIMP is a Majorana fermion. The Lagrangian for the second model is,

\begin{equation}
L = \frac{1}{2} \bar{\chi} \cancel{\partial} \chi - \frac{m_{\chi}}{2} \bar{\chi}\chi
\end{equation}

\noindent
As outlined in Ref \cite{Bae:2001id}, in the Randall-Sundrum model, the radion couples to the trace of the energy-momentum tensor, denoted by $\Theta_{\mu}^{\mu} $, 

\begin{equation}
L_{int} = \frac{\phi}{\Lambda_{\phi}} \Theta_{\mu}^{\mu} 
\end{equation}

\noindent
where $\Lambda_{\phi}$ is the vacuum expectation value of the radion. The couplings of the radion to the Standard Model fields was derived in Ref \cite{Bae:2001id}, and for the case of strongly warped extra dimensions are similar to the Higgs couplings.

\par
It should be noted that in the figures for this model, it is assumed that $\Lambda_{\phi} = v_{EW}$. While solving the hierarchy problem does require the size of the extra dimensions to be stabilized with $\Lambda_{\phi} \sim O(TeV)$ \cite{Goldberger:1999un}, there is no further restriction on its size. For comparison with the previous models which rely on a Higgs coupling, and following the examples in Ref \cite{Bae:2001id}, it will be assumed that $\Lambda_{\phi} = v_{EW}$ for the purpose of each calculation. The actual $\Lambda_{\phi}$ dependence included in an effective coupling constant,

\begin{equation}
\label{Eq:Kappa}
\kappa \equiv \left( \frac{m_{S,f}}{1 \; TeV}^2 \right)\left( \frac{v_{EW}}{\Lambda_{\phi}} \right)^2 \left(  \frac{1\; TeV}{M_{\phi}}\right)^2
\end{equation} 

\noindent
where $M_{\phi}$ is the mass of the radion. It should also be noted that in the range of $m_{S,f} \gg \Lambda_{\phi}$ the couplings can become non-perturbative and therefore such heavy WIMPs are not considered in this model.

\section{Abundance Constraints \label{Section:RSDM}}

\par
As outlined in Ref. \cite{Kolb:1990vq}, the cosmic abundance of dark matter is determined by the thermally averaged annihilation cross-section.
The particles in this model annihilate via a virtual radion, which subsequently decays into Standard Model fields. The annihilation cross-section can be written in terms of the radion decay width, given in Ref \cite{Bae:2001id},

\begin{equation}
<\sigma_{s} v> = \frac{8 M_S^4}{\Lambda_{\phi}^2} \frac{1}{(4 M_S^2 - M_{\phi}^2)^2 + M_{\phi^2}\Gamma_{\phi}^2}  \left( \frac{\Gamma_{\phi \to X}}{M_{\phi}} \right)_{M_{\phi} \to 2 M_S}
\end{equation}

\begin{equation}
<\sigma_{f} v> = \frac{12 m_f^3 T}{\Lambda_{\phi}^2} \frac{1}{(4 m_f^2 - M_{\phi}^2)^2 + M_{\phi}^2\Gamma_{\phi}^2}  \left( \frac{\Gamma_{\phi \to X}}{M_{\phi}} \right)_{M_{\phi} \to 2 m_f}
\end{equation}



\begin{figure}
\begin{center}
\psfig{file=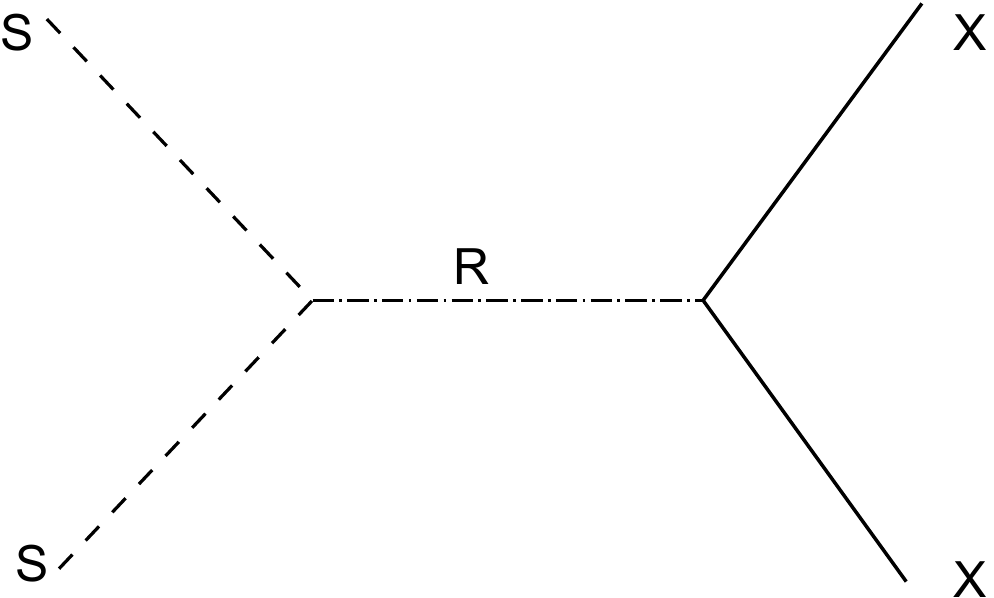,width=0.65\textwidth,angle=0}
\end{center}
\caption{\label{Figure:EXDM-ann} Feynman diagram for the annihilation cross section of WIMPs in the presence of warped extra dimensions. In this diagram, R represents the radion which acts as a mediator for the annihilation and S represents either scalar or fermionic WIMPs.}
\end{figure}

\noindent
where the first equation corresponds to scalar WIMPs and the second to fermionic WIMPs, and $\Lambda_{\phi}$ is the vacuum expectation for the radion field. As discussed in Ref \cite{Griest:1990kh}, this form of the thermally average cross-sections is only valid when the WIMP mass is not close to the resonance in the radion propagator, and not close to a threshold for producing heavier Standard Model fields. For these regions the thermal average of the cross-sections have been calculated numerically.

\begin{figure}
\psfig{file=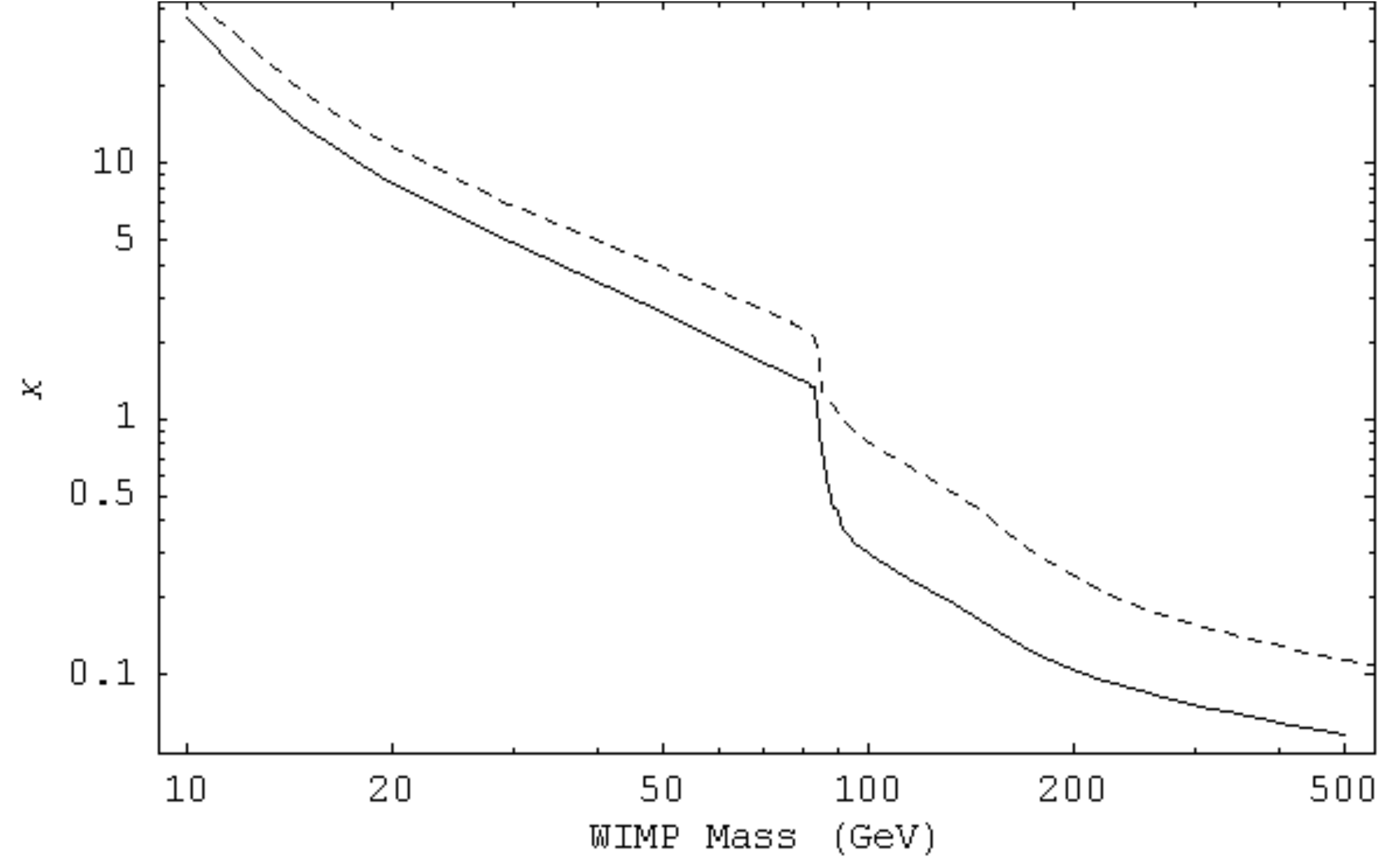,width=\textwidth,angle=0}
\caption{\label{Figure:RS-AC} Abundance constraints on scalar (solid line) and fermionic (dashed line) WIMPs in the presence of warped extra dimensions. 
}
\end{figure}

\par
The dark matter abundance is then calculated using standard methods (see for example Ref \cite{Kolb:1990vq}),

\begin{equation}\label{Eq::Abundance}
\Omega_{DM} h^2 = \frac{1.07 \times 10^9 x_f}{g_*^{1/2} M_{PL} \; GeV \; <\sigma_{ann} v>}
\end{equation}

\noindent
and the results are plotted in Figure \ref{Figure:RS-AC} in terms of the effective coupling constant given in Eq \ref{Eq:Kappa}.

\par
By requiring $\kappa$ to be perturbative, which in this article will be assumed to mean $\kappa \lesssim \sqrt{4 \pi}$, an additional constraint is imposed on the model (referred to as the {\it Lee-Weinberg} limit\cite{Lee:1977ua}). From the abundance constraints given above, this bound requires $m_S \gtrsim 35 \;GeV$ and $m_f \gtrsim 50 \; GeV$. For both scalars and fermions there is a lowering of $\kappa$ at $m_{S,f} \sim 85 \;GeV$, due to the availability of annihilations to gauge bosons. This decay channel is efficient, leading to a larger cross-section and requires smaller values of $\kappa$ to produce the correct dark matter abundance. It should also be noted that the coupling of the WIMPs to the radion is determined by the mass of the WIMP, and therefore the abundance constraints leave only the radion mass as a free parameter.  

\par
The calculations and results of this section demonstrate that the presence of warped extra dimensions can allow a WIMP to have no gauge or Yukawa interactions with other particles, but still annihilate efficiently through gravitational forces to provide the correct dark matter abundance. 

\section{Dedicated Dark Matter Searches}

\par
At present, the primary method of searching for dark matter is through dedicated dark matter experiments in which WIMPs in the solar system collide with the nuclei in the detectors, and effects of the recoil detected. Such experiments impose an upper limit on the WIMP-nucleon scattering cross-section.

\par
The scattering cross-section for the models in this article depend on the radion-nucleon coupling. This coupling is given by the trace of the energy momentum tensor of the nucleon,

\begin{equation}
L_{int} = \frac{\phi}{\Lambda_{\phi}} \Theta_{\mu}^{\mu}
\end{equation}

\noindent
where, at low energies,

\begin{equation}
<N| \Theta_{\mu}^{\mu} |N> = m_N <N| \bar{\psi}_N \psi_N |N>
\end{equation}

\noindent
This coupling is stronger than the corresponding Standard Model Higgs coupling, as the radion couples to the full energy-momentum tensor rather than just to the mass of the quarks and quark loops in the nucleon. This difference enhances the WIMP-nucleon scattering cross-section in this model by an order of magnitude compared to the Minimal Model of Dark Matter.

\begin{figure}
\psfig{file=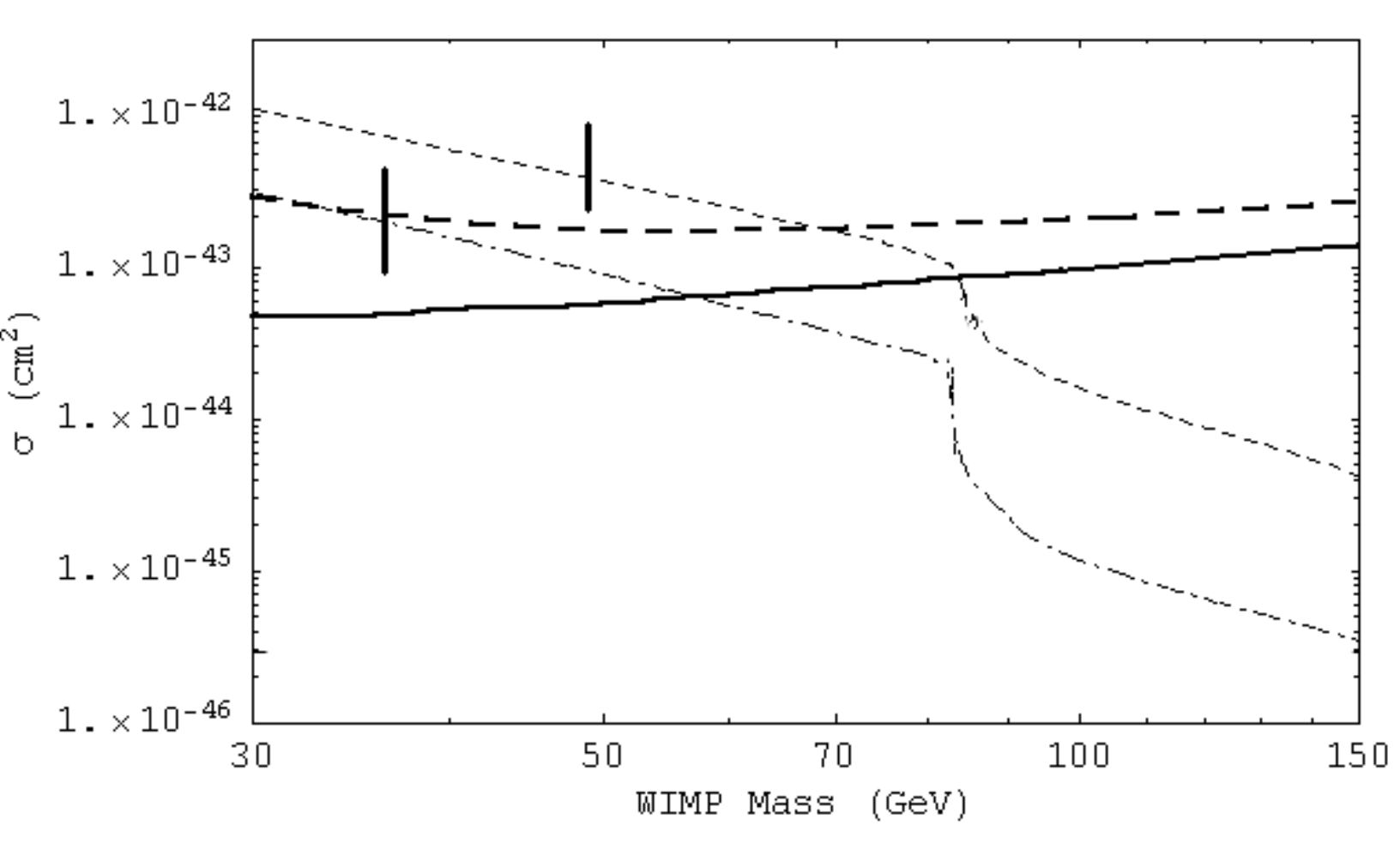,width=\textwidth,angle=0}
\caption{\label{Figure:RS-NC} WIMP-nucleon scattering cross-section for scalar WIMPs (dashed-dotted line) and fermion WIMPs (dashed line) with radion mediation. Limits from CDMS and XENON10 are indicated with the bold dashed and solid lines respectively. The short vertical lines represent the Lee-Weinberg bound on scalar and fermion WIMPs.}
\end{figure}

\par
For the scalar WIMPs, the WIMP-nucleon scattering cross section is

\begin{equation}
\sigma_{S} = \frac{m_N^4 m_S^2}{2 \pi \Lambda_{\phi}^4 m_{\phi}^4} = 1.3 \times 10^{-44} cm^2 \kappa^2
\end{equation}

\noindent
while for fermions the scattering cross-section is

\begin{equation}
\sigma_{f} = \frac{m_N^4 m_f^2}{\pi \Lambda_{\phi}^4 m_{\phi}^4} = 2.6 \times 10^{-44} cm^2 \kappa^2
\end{equation}

\noindent
Using the abundance constraint previously derived, the scattering cross-sections can be calculated, and the results are plotted in Figure \ref{Figure:RS-NC} along with constraints from the CDMS\cite{Akerib:2005kh} and XENON\cite{Angle:2007uj} experiments\footnote{During the preparation of this article, new data was released by the CDMS experiment\cite{Ahmed:2008eu}. However the updated data does not improve the constraints on this model, as the region of parameter space which it excludes is already excluded by other experiments}.  

\par
The reported bounds from the XENON10\cite{Angle:2007uj} experiment provides the strongest constraint, with scalars excluded for $m_{S} \lesssim 57 \; GeV$ and fermions excluded for $m_f \lesssim 85 \; GeV$. 

\section{Collider Constraints}

\par
Another method of searching for WIMPs in this model is to search for invisible radion decays at the LHC or other high energy colliders. Due to the similarities of the radion and the Higgs couplings, this signal is similar to the signal for an invisible Higgs boson with the primary channels being $pp \to Z + R \to Z + DM$ and production of radions through weak-boson fusion. Both of these channels and their backgrounds have been extensively studied for the Higgs boson \cite{Martin:1999qf,Eboli:2000ze,Davoudiasl:2004aj,Zhu:2005hv}, and the results are expected to be very similar for radions. The primary difference between the production of Higgs bosons and the production of radions is in rate of production via gluon fusion. However as outlined in \cite{Davoudiasl:2004aj} this channel is obscured by a large background signal and will not significantly improve on the signal for either the Higgs boson or the radion.

\begin{figure}
\psfig{file=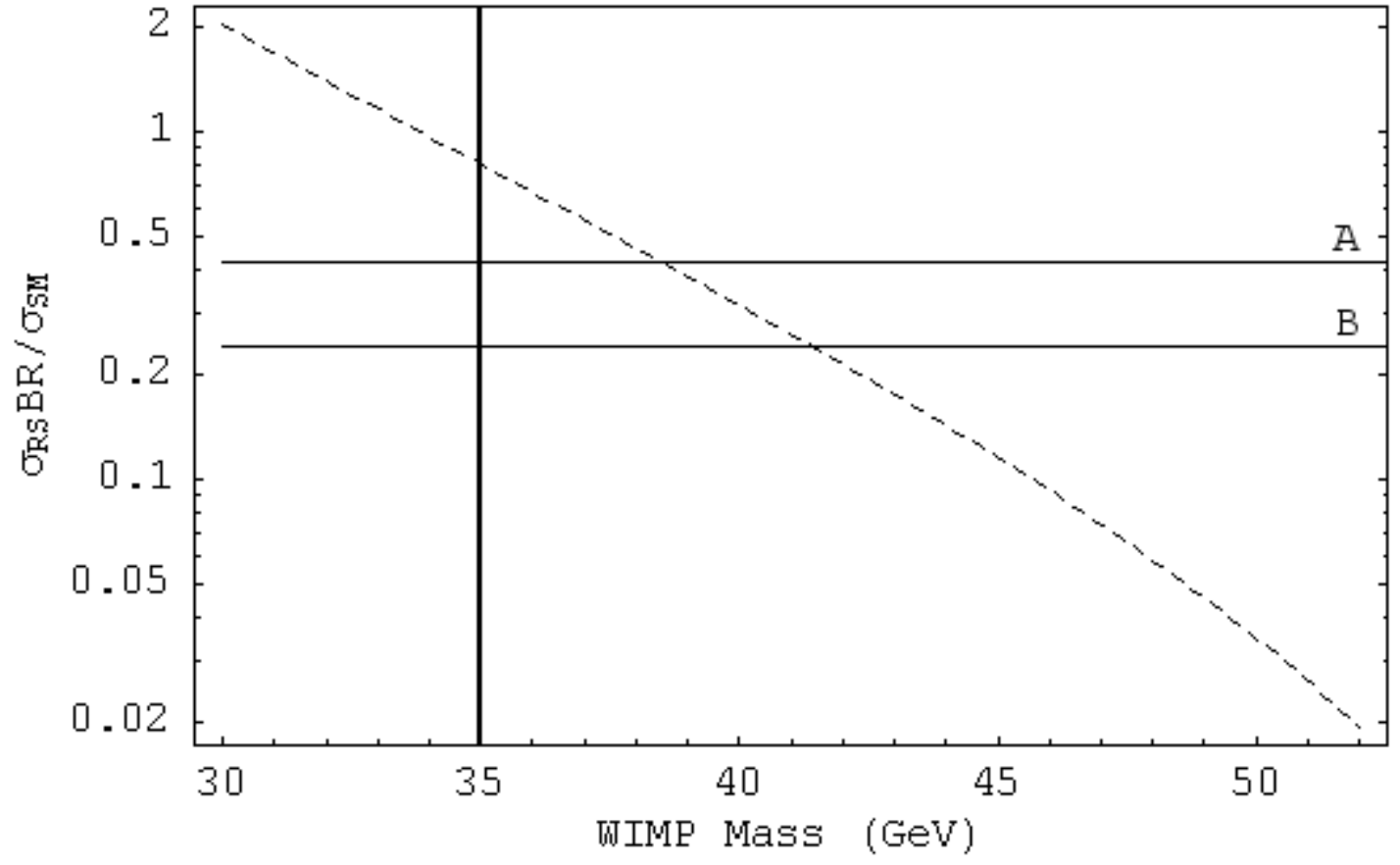,width=\textwidth,angle=0}
\caption{\label{Figure:RS-collider} Sensitivity of the LHC to scalar WIMPs through the invisible radion signal. Lines A and B represent the smallest branching ratio that can be detected at ${\mathcal L} = 10 \; fb^{-1}$ and ${\mathcal L} = 30 \; fb^{-1}$ respectively. The vertical line represents the WIMP mass at which the model becomes non-perturbative.}
\end{figure}

\par
The most significant difference in the production cross-sections is that, unlike in the Higgs models in which $v_{EW}$ is fixed and therefore the Standard Model couplings are known, the radion vev $\Lambda_{\phi}$ is unknown. As a result, the cross-sections for radion production are enhanced (or suppressed) by a factor of $\sigma_{RS}/\sigma_{SM} \approx v_{EW}^2/\Lambda_{\phi}^2$ relative to the production cross-section for Higgs bosons. 

\par
As is common in calculations of experimental sensitivity to invisible Higgs boson decays, the sensitivity of experiments to invisible radion decays is expressed in terms of the relevant branching ratio of the radion. The branching ratio for the decay of a radion to scalar WIMPs in this model is plotted in Figure \ref{Figure:RS-collider}, with a scaling factor added to account for the $\Lambda_{\phi}/v_{ew}$ enhancement in the radion production cross-section. The range of branching ratios which can be probed at the $3\sigma$ level by the LHC for luminosities ${\mathcal L} = 10 \; fb^{-1}$ and ${\mathcal L}= 30 \; fb^{-1}$ are also given. For the purpose of comparison with previously published models, it is assumed that $M_{\phi} = 120 \; GeV$ while $m_S$ and $\Lambda_{\phi}$ are varied to satisfy the abundance constraints. In addition, only scalar WIMPs are considered here, as the fermionic model becomes non-perturbative for the regions of parameter space which can be effectively probed by the LHC or Tevatron. 

\par
Using these results, the LHC can probe scalar WIMPs in this model up to $m_S \lesssim 38 \;GeV$ with ${\mathcal L} = 10 \; fb^{-1}$ and up to $m_S \lesssim 42 \;GeV$ with ${\mathcal L} = 30 \; fb^{-1}$, while perturbative couplings require $m_S \gtrsim 35 \;GeV$. However as demonstrated in the previous section, this range of WIMP masses has already been excluded by dedicated dark matter searches\footnote{It may still be possible to probe both the scalar and fermionic WIMP models in the LHC or Tevatron if the radion mass is significantly heavier, however a detailed calculation of the invisible radion production cross-section for heavier radions is beyond the scope of this paper.}.

\section{Summary}

\par
The two models presented in this article are minimal models, in that they involve a single field to be added to the Standard model and the presence of a single extra dimension. In contrast to previously published minimal models, these dark matter candidates do not require any additional interactions to be added to the Lagrangian.

\par
From perturbative requirements and abundance constraints, the mass of the new particle must be $m_S \gtrsim 35 \; GeV$ for scalar WIMPs and $m_f \gtrsim 50 \; GeV$ for fermions, while experimental constraints from nuclear recoil experiments can further exclude masses as high as $m_S \sim 60 \; GeV$ for scalar WIMPs and $m_f \sim 80 \; GeV$ for fermions. Depending on the mass of the radion in this model, the LHC and other future colliders may be able to probe higher WIMP masses through the invisible radion signal.

\par
In conclusion, it has been demonstrated in this article that there can exist a candidate for dark matter which requires a minimum of new physics, evades existing experimental constraints, while also predicting the correct abundance of dark matter, but which interacts only through gravitational forces and thus requires no additional interactions. 

\bibliography{rmdm}

\end{document}